\documentclass[prl,amsmath,amssymb,aps,superscriptaddress,floatfix,reprint,notitlepage]{revtex4-1}
\usepackage{times}

\usepackage{amsmath, braket, amsfonts}
\usepackage{amssymb}
\usepackage{natbib}
\usepackage{sidecap}
\usepackage{bm, color, ulem}
\usepackage{xcolor}
\usepackage{lipsum}

\usepackage{graphicx}
\usepackage{dcolumn}
\usepackage{bm}

\begin{document}
\setcitestyle{super}
\preprint{APS/123-QED}

\title{Harnessing excitons at the nanoscale - photoelectrical platform for quantitative sensing and imaging}
\author{Zhurun Ji}
\affiliation{Department of Physics, Stanford University, Stanford, 94305, CA, USA}
\affiliation{Department of Applied Physics, Stanford University, Stanford, 94305, CA, USA}
\affiliation{Geballe Laboratory for Advanced Materials, Stanford University, Stanford, 94305, CA, USA}
\author{Mark E. Barber}
\affiliation{Department of Physics, Stanford University, Stanford, 94305, CA, USA}
\affiliation{Department of Applied Physics, Stanford University, Stanford, 94305, CA, USA}
\affiliation{Geballe Laboratory for Advanced Materials, Stanford University, Stanford, 94305, CA, USA}
\author{Ziyan Zhu}
\affiliation{Stanford Institute for Materials and Energy Sciences, SLAC National Accelerator Laboratory, Menlo Park, 94025, CA,USA}
\author{Carlos R. Kometter}
\affiliation{Department of Physics, Stanford University, Stanford, 94305, CA, USA}
\affiliation{Geballe Laboratory for Advanced Materials, Stanford University, Stanford, 94305, CA, USA}
\affiliation{Stanford Institute for Materials and Energy Sciences, SLAC National Accelerator Laboratory, Menlo Park, 94025, CA,USA}
\author{Jiachen Yu}
\affiliation{Department of Applied Physics, Stanford University, Stanford, 94305, CA, USA}
\affiliation{Geballe Laboratory for Advanced Materials, Stanford University, Stanford, 94305, CA, USA}
\affiliation{Stanford Institute for Materials and Energy Sciences, SLAC National Accelerator Laboratory, Menlo Park, 94025, CA,USA}
\author{Kenji Watanabe}
\affiliation{Research Center for Electronic and Optical Materials, National Institute for Materials Science, 1-1 Namiki, Tsukuba, 305-0044Japan}
\author{Takashi Taniguchi}
\affiliation{Research Center for Materials Nanoarchitectonics, National Institute for Materials Science, 1-1 Namiki, Tsukuba, 305-0044, Japan}
\author{Mengkun Liu}
\affiliation{Department of Physics and Astronomy, Stony Brook University, Stony Brook, 11794, NY, USA}
\affiliation{National Synchrotron Light Source II, Brookhaven National Laboratory, Upton, 11973, NY, USA}
\author{Thomas P. Devereaux}
\affiliation{Geballe Laboratory for Advanced Materials, Stanford University, Stanford, 94305, CA, USA}
\affiliation{Stanford Institute for Materials and Energy Sciences, SLAC National Accelerator Laboratory, Menlo Park, 94025, CA,USA}
\affiliation{Department of Materials Science and Engineering, Stanford University, CA 94305, USA}
\author{Benjamin E. Feldman}
\affiliation{Department of Physics, Stanford University, Stanford, 94305, CA, USA}
\affiliation{Geballe Laboratory for Advanced Materials, Stanford University, Stanford, 94305, CA, USA}
\affiliation{Stanford Institute for Materials and Energy Sciences, SLAC National Accelerator Laboratory, Menlo Park, 94025, CA,USA}
\author{Zhixun Shen}
 \email{zxshen@stanford.edu}
\affiliation{Department of Physics, Stanford University, Stanford, 94305, CA, USA}
\affiliation{Department of Applied Physics, Stanford University, Stanford, 94305, CA, USA}
\affiliation{Geballe Laboratory for Advanced Materials, Stanford University, Stanford, 94305, CA, USA}
\affiliation{Stanford Institute for Materials and Energy Sciences, SLAC National Accelerator Laboratory, Menlo Park, 94025, CA,USA}


\begin{abstract}
\textbf{
Excitons --- quasiparticles formed by the binding of an electron and a hole through electrostatic attraction --- hold promise in the fields of quantum light confinement and optoelectronic sensing. Atomically thin transition metal dichalcogenides (TMDs) provide a versatile platform for hosting and manipulating excitons, given their robust Coulomb interactions and exceptional sensitivity to dielectric environments. In this study, we introduce a cryogenic scanning probe photoelectrical sensing platform, termed exciton-resonant microwave impedance microscopy (ER-MIM). ER-MIM enables ultra-sensitive probing of exciton polarons and their Rydberg states at the nanoscale. Utilizing this technique, we explore the interplay between excitons and material properties, including carrier density, in-plane electric field, and dielectric screening. Furthermore, we employ deep learning for automated data analysis and quantitative extraction of electrical information, unveiling the potential of exciton-assisted nano-electrometry. Our findings establish an invaluable sensing platform and readout mechanism, advancing our understanding of exciton excitations and their applications in the quantum realm.}
\end{abstract}

\maketitle

\onecolumngrid
One of the fundamental tenets of quantum mechanics is the existence of discrete energy levels, which underpins various applications in the field of quantum technology. For instance, qubits, the building blocks of quantum computing and communication, leverage this concept to perform complex calculations\cite{saffman2010quantum}. Quantum sensing techniques that make use of discrete energy levels, such as nitrogen-vacancy and Rydberg atom sensing, enable precise detection of electric and magnetic field\cite{casola2018probing,schirhagl2014nitrogen,degen2017quantum}. The success of these techniques and the aspiration to probe additional physical quantities at the nanoscale motivate the development of novel discrete energy systems and sensing mechanisms.

While most of the collective excitations in quantum materials form continuous energy bands, there are certain types of discretized energy states due to quantum confinement, disruption of crystalline order or strong magnetic fields. One type of collective excitation, exciton, has discrete energy levels from the Coulomb attraction between the negatively charged electron and the positively charged hole. Their energy structure, coherence, and interactions with photons make excitons an ideal candidate for sensing devices\cite{popert2022optical,chand2022visualization,xu2020correlated}. Atomically thin transition metal dichalcogenides (TMDs) are particularly attractive for hosting robust excitonic effects, stemming from their large binding energies, direct bandgap, and pronounced light-matter interactions \cite{manzeli20172d,wang2018colloquium}. The intrinsic Coulomb interactions within excitons give rise to strong exciton-exciton and exciton-phonon coupling, culminating in intriguing phenomena such as exciton condensation and the formation of, e,g., trions\cite{lui2014trion} and exciton-polarons\cite{efimkin2021electron}. Under an electric field, these interactions can substantially alter exciton properties, impacting quantities like binding energies, radiative lifetimes, and emission spectra\cite{klein2016stark}. The rich response of excitons to external stimuli suggests the potential for effective control and sensing. Importantly, in compact two-dimensional (2D) heterostructures, excitons provide the capability for high sensing accuracy, as they can be positioned in close proximity to the samples\cite{mak2022semiconductor,regan2022emerging,cai2023signatures,regan2020mott}. 

To successfully harness excitons as nanoscale sensors, the development of an excitation and imaging method is imperative. Detecting excitons locally is a critical first step, allowing one to gather precise data without disturbing the quantum mechanical properties. While tip-enhanced photoluminescence (TEPL) techniques have been used to study nanoscale excited state recombination in TMDs at room temperature \cite{bao2015visualizing,darlington2020imaging,chand2022visualization,zhou2023near,hasz2022tip}, and scattering-type scanning near-field optical microscopes (s-SNOM) have provided insights into the complex dielectric function of excitons \cite{plankl2021subcycle,zhang2022nano} and their waveguide modes \cite{hu2017imaging,luan2022imaging,iyer2022nano}, these methods face limitations. 
Precise sensing requires a narrow exciton absorption linewidth, typically achieved at lower temperatures with direct optical excitation. However, both methods become challenging at extremely low temperatures (i.e., below 4 K), and spectroscopic measurements of excitons have only been demonstrated at room temperature so far. Additionally, emission experiments are susceptible to quenching due to environmental factors, complicating the analysis of exciton-environment interactions. These multifaceted challenges underscore the pressing need for an innovative hyperspectral imaging technique that can fully capture the fine spectroscopic details of excitons at the nanoscale in a cryogenic environment. 

In this study, we develop an advanced technique, exciton resonant microwave impedance microscopy (ER-MIM), to visualize excitons with nanoscale precision at extremely low cryogenic temperatures. By combining the spectral resolution of optical spectroscopy with the high spatial resolution and enhanced sensitivity to complex dielectric response at microwave regime of MIM, we gain deep insights into exciton dynamics and their interactions with the local environment. This simultaneous imaging of dielectric responses and spectroscopic measurements of exciton energies at nano-scale enable the implementation of machine learning algorithms to reconstruct the all three important electrical quantities --- conductivity, electric field and surrounding permittivity. This ``all-in-one" capability for local electrical sensing paves the way for advances in quantum material-based sensing.

\section{Results}\label{sec2}
\subsection{Local optoelectronic detection of discrete exciton Rydberg states}
We first introduce the ER-MIM technique that enables us to sense excitons using photoelectrical effects at the nanoscale. As shown in Fig. 1a, our setup consists of two parts: a microwave transmission line impedance-matched to a metallic scanning probe (tip) \cite{chu2020unveiling,barber2022microwave,cui2016quartz},  and a continuously wavelength-tunable laser that is fiber-coupled into a Helium-3 cryostat and illuminates the tip. The real and imaginary parts of the reflected microwave signals are recorded, which are in- and out-of-phase with the reference microwave excitation line, respectively. Our measurements were conducted at $\sim$1.5 K and 3 GHz, yet this method can readily be adapted to even lower temperatures (e.g., the base temperature of the Helium-3 cryostat). In Fig. 1b, we present two calculated response curves that illustrate the relationship between the real (in-phase) and imaginary (out-of-phase) parts of the reflected microwave signals and the local complex conductivity of the sample.
 
The material we study is a prototypical 2D TMD device — a monolayer of MoSe$_2$ encapsulated within hBN layers with graphite as a back gate. The optical spectrum of monolayer MoSe$_2$ possesses a series of pronounced resonances, categorized as A- and B-excitons (see Fig. 1c). These excitons arise from two optical transitions involving states in the upper and lower energy spin valence bands. They possess Rydberg atom-like energy levels, each designated by its principal quantum number, $n$.

We track the evolution of the microwave response as a function of excitation wavelength and back gate voltage $V_\text{bg}$ controlling the free carrier density in MoSe$_2$ (imaginary part of the ER-MIM signal shown in Fig. 1d). From the response curve in Fig. 1b, the blue and red colored regions correspond to low and high conductivity of MoSe$_2$, respectively. The overall dependence of conductivity on back gate voltage agrees with the transfer curve of a semiconductor, i.e., it decreases when crossing from the conduction band into the bandgap, and increases when entering the valence band. On top of this overall trend, at specific photon energies (e.g., $~$1.65 eV), the conductivity has an extra increase at smaller $|V_\text{bg}|$ (e.g., $<$ 2V), and another decrease at large $|V_\text{bg}|$ (e.g., $>$ 2V). We identify those photon energies as in resonance with neutral $A_\text{1s}$ and $B_\text{1s}$ excitons, and the excited Rydberg states including $A_\text{2s}$, $A_\text{3s}$, and $B_\text{2s}$, as well as weaker modulations in intensity features corresponding to their charged counterparts like $AT_\text{1s}$ and $AT_\text{2s}$ (which are usually referred to as trion or exciton polaron, and their distinction will be discussed below).

The changes in ER-MIM-Im signal near excitonic resonances imply the observation of photoconductivity associated with exciton formation. The sign flip from positive (red) at small $|V_\text{bg}|$ to negative (blue) at large $|V_\text{bg}|$ can be attributed to two major processes (a discussion on ruling out other possible optical processes in note 2 \cite{supp}). One is the photoconductive effect \cite{chu2020unveiling}, where the exposure of TMD-based devices to light generates neutral electron-hole pairs. Then, the trapping of one carrier and the release of the other free carrier from an exciton could result in an enhanced conductivity \cite{handa2023spontaneous}. This process explains our observation of positive conductivity near excitonic resonances. The other is the Auger assisted tunneling effect\cite{chow2020monolayer,sushko2020asymmetric}. When the photon energy is in resonance with the binding energy of excitons, a generated exciton can excite a free hole when it recombines during the Auger process. The hole can subsequently tunnel through the hBN barrier to reach the bottom gate, which effectively reduces the MoSe$_2$ conductivity. Since the energy barrier on the hole side between MoSe$_2$ and hBN is much smaller than the electron side, the negative photoconductivity is only observed when MoSe$_2$ is hole-doped \cite{supp}. Therefore, the conductivity decrease at large $|V_\text{bg}|$ is attributed to the dominance of the Auger effect. 

We further compare two linecuts of ER-MIM-Im spectrum with previously reported far-field spectra of excitons in monolayer MoSe$_2$ (Fig. 1e). The red and blue curves show excitonic peaks and dips, highlighting the signal from the photoconductivity and Auger-assisted tunneling effects, respectively. The comparison reveals excellent agreement between the identified exciton energies in ER-MIM-Im and the results obtained from optical reflectance\cite{arora2015exciton}, photoluminescence\cite{liu2021exciton}, and photocurrent\cite{vaquero2022low} measurements. It is noteworthy that our spectral resolution stands on par with, or surpasses, those of the above-mentioned area-averaged methods. This compelling correspondence demonstrates that our ER-MIM technique enables accurate photoelectric measurements of exciton spectra with the added benefit of exceptional spatial resolution and access to a temperature regime that has, until now, been unattainable.

\subsection{The interplay between excitons and the charge environment}

~~~~~To untangle the exciton-electron interactions at the nanoscale, we modulate carrier density of MoSe$_2$ by controlling $V_\text{bg}$ applied onto graphite while collecting ER-MIM spectra (Fig. 1a). The graphite boundary splits MoSe$_2$ into three distinct spatial regions (Fig. 2a): (I) a region that extends beyond the graphite back gate (BG) that is ungated, (II) the region directly above the BG, in which the carrier density is controlled by $V_\text{bg}$, and (III) the depletion region associated with the nanojunction between regions I and II. Our focus lies on contrasting the ER-MIM responses in regions I and II, which are distant from the junction and hence less influenced by the electric field, yet differentiated by their carrier densities. Such a comparative approach enables a systematic exploration of excitonic behaviors in diverse local charge environments.

Fig. 2b depicts two scenarios of exciton-electron interactions. The first involves trions, charged and weakly coupled three-particle complexes formed by binding two electrons (or holes) to one hole (or electron). Trion and exciton eigenenergies differ by the trion binding energy, denoted as $\rm E_T$ (left schematic). The second scenario features charge-neutral bosons resulting from excitons interacting with the degenerate Fermi sea of excess charge carriers (right schematic). In this context, excitons become dressed by Fermi sea excitations, giving rise to attractive and repulsive exciton-polaron (AP and RP) quasiparticles, akin to Fermi-polarons in the context of cold atoms \cite{liu2021exciton, efimkin2021electron}. The energy difference between AP and RP branches, or the exciton polaron binding energy is approximately $\rm E_T +3/2\epsilon_F$, with $\epsilon_F$ representing the Fermi energy \cite{efimkin2021electron}. To examine these two physical pictures, we take advantage of the sensitivity of ER-MIM spectra to $\epsilon_F$ (gate)-tunable local excitonic energy levels.

In the upper panel of Fig. 2c, we present the ER-MIM spectrum obtained in region I. This region is not strongly affected by the back gate voltage, hence close to intrinsically ($n$-)doped. The spectrum reveals one major resonance feature at $\sim$1.642~eV, corresponding to exciton $A_\text{1s}$ (fitting shown as a grey dashed line). Notably, at 1.5~K, the linewidth of the exciton $A_\text{1s}$ dip is $\rm \sim$2.5~meV, closely approaching the theoretically calculated intrinsic linewidth of excitons \cite{selig2016excitonic}. Similar linewidth was observed in the spectrum measured in the more $p$-doped region III (shown in the lower panel of Fig. 2c). The spectrum shows one more prominent feature centered at $\sim$1.59~eV, corresponding to trion $T_\text{1s}$ (fitting shown as a grey dotted line). 

On the other hand, we conducted spatially-resolved ER-MIM measurements at four distinct locations spaced 50 nm apart in $p$-doped region II (Fig. 2a, spots 1-4) at a constant $\rm V_{bg}=-2.6\,V$. The line spectra zoomed in near the trion resonance energy are shown in Fig. 2d. This region has the highest carrier density among the three, and the formerly designated $T_\text{1s}$ in region III feature divides into two separated dips (marked by the red dashed and blue dotted lines). Moreover, the intensity of the two dips exhibits significant variations across different spots.

To further understand the spatial dependence, we measured the back gate dependence of the ER-MIM spectra sitting at spot 4. The red and blue circles in Fig. 2e are extracted exciton eigenenergies at different $V_\text{bg}$ near formerly designated $T_\text{1s}$, and the green circles in Fig. 2f are another eigenenergy near formerly designated $A_\text{1s}$. The red dashed and blue dotted $T_\text{1s}$ branches display nearly linear redshifts while the $A_\text{1s}$ exciton energy demonstrates a linear blueshift with $V_\text{bg}$. This trend remains consistent throughout the range of $V_\text{bg}$ in our experiments. The increasing energy separation between the branches in Fig. 2e and the other one in Fig. 2f is a signature of exciton-polaron formation at elevated carrier doping levels (Fig. 2b). In this framework, the formerly assigned $A_\text{1s}$ becomes repulsive exciton-polaron (RP). The $T_\text{1s}$ branch turns into attractive exciton-polarons (AP$_1$ and AP$_2$), where the splitting between AP$_1$ and AP$_2$, a feature obscured in far field experiments but restored in ER-MIM signal, can be understood as excitons coupling with the Fermi seas in two different valleys (Fig. 2b). 

The observed gate-dependent shifts in AP and RP eigenenergies indicate interactions between excitons and free carriers. These energy shifts exhibit an almost linear dependence on $V_\text{bg}$, reflecting a collective effect of carrier density dependent exciton binding energy change, quasiparticle bandgap renormalization, and the variation of binding energy of the exciton polaron. To quantify the shifts, we introduce a coefficient $\beta$, where $\delta E = \beta n$, with $\delta E$ being the change of eigenenergy of an exciton polaron state which has a linear dependence on carrier density $n$. The observed nanoscale spatial variation of $\delta E$ suggests that the eigenenergy changes can be used as a sensitive local indicator of the charge environment.

\subsection{The interplay between excitons and the electric environment}
In addition to the charge environment, to understand the intricate interplay between excitons and the local electrical variations, we examine more closely the ER-MIM signals in region III, the depletion area. Here, an in-plane electric field propels electrons and holes towards regions I and II, respectively. We also compare the the excitonic response closeby and farther away from the controlled $p$-$n$ junction, i.e., region III and region II, as illustrated in the schematic in Fig. 3a.

To start with, a systematic ER-MIM scan across the $p$-$n$ junction was carried out as a function of $V_\text{bg}$ and excitation energy. We mark the locations in Fig. 3b and plot the results in Fig. 3c. It is evident that a continuous variation in the ER-MIM-Im signal near the $\rm A_{1s}$ AP and RP resonances is observed across the $p$-$n$ junction. Especially, when on resonance, there is a continuous change from positive to negative photoconductivity from the left-hand side of the nanojunction to the right-hand side. This marks the conversion from a photoconductive effect-dominated mechanism in the ungated region of MoSe$_2$ to Auger tunneling-dominated mechanism in the back-gated region of MoSe$_2$. 

The field effect is then studied in greater detail via high-resolution imaging in region III (Fig. 3d and 3e, with locations marked by stars in Fig. 3b). The two gate-dependent spectra in region III both appear to be very different from the ones obtained from region II (Fig. 1b). They feature mostly positive photoconductivity at excitonic resonances (red colored regions), and less change with gating. Within region III, measurement in the right part (purple star in Fig. 3b, Fig. 3e) shows a relatively more effective gating effect, where the restoration of some negative photoconductivity can be identified. This observation further elucidates the $p$-$n$ junction behavior in region III.

To study the excitonic response to electric field, we extracted the RP eigenenergy from Fig.~3e, and plot it in Fig.~3f. There are two distinct features that differ from region II (Fig.~2f): the RP energy exhibits an overall redshift behavior with doping level, instead of a blueshift; the RP energy has an abrupt blueshift at $\rm V_{bg} \sim -2.2$~V. The first feature agrees with the dc Stark effect of excitons: $\delta E= -\alpha F^2/2$, where $\alpha$ is the exciton polarizability \cite{klein2016stark}, $F$ represents the electric field strength and $\delta E$ signifies the energy shift. The second feature underscores the nonlocal effect of the dielectric constant on exciton energies. Since a carrier density gradient exists in this region, when the right part of the sample goes across the band edge, it leads to a sharp increase in carrier compressibility $\frac{1}{n^2}(\frac{\partial n}{\partial \mu})$ (n being carrier density, and $\mu$ being the chemical potential), which then changes the exciton screening in the left part. This is consistent with the Coulomb potential screening of excitons through a free electron gas, which leads to another describable dependency of excitons on the environmental permittivity: $E = E(\kappa)$. These observations suggest how excitons can serve as imaging tools for a $p$-$n$ junction and reveal intricate nanoscale parameters such as electric field and dielectric constant.

\subsection{Quantitative photoelectrical imaging and exciton-facilitated electrometry}
To quantitatively analyze the ER-MIM data and utilize it for predicting material properties, we developed a deep neural network (DNN) \cite{kalinin2015big,krull2020artificial,kalinin2021automated,chen2023machine,kelley2020fast} based on a set of MIM maps. These maps consist of 256$\times$100 pixel images, obtained with variations in optical excitation energies and back gate voltages within a 1~$\mathrm{\mu}$m $\times$ 1~$\mathrm{\mu}$m region at the graphite back gate boundary. Due to the complex, multidimensional nature of the datasets, traditional fitting approaches prove to be inadequate. In Fig. 4a, we present three representative sets of MIM-Im and MIM-Re data collected at $\rm V_{bg}= -3~V$, under different conditions: with the light off, under 1.651 eV light excitation, and 1.655 eV light excitation (complete dataset in \cite{supp}). The 1.651 eV excitation is off-resonance with excitons, while the 1.655 eV excitation is close to the resonance of RP branch of the $A_\text{1s}$ exciton.

In the following, we show how to use this DNN to predict the local environment, ($\sigma$, $\kappa$, $h$), i.e., sheet conductivity of MoSe$_2$, the relative permittivity of the hBN-MoSe$_2$-hBN stack, and relative tip-sample distance $h$, for a given MIM data set. We construct a feedforward network (Fig.~4b), with the inputs being MIM-Re and MIM-Im, either from the experiment or finite element simulations\cite{multiphysics1998introduction,supp}, and predict the probability at each set of discretized $(\sigma, \kappa, h)$ values. 
We train the network with a total of 64,000 sets of simulation data, which is divided into bins along the $\sigma$, $\kappa$, and $h$ directions.
Each data point is encoded into a 3840-dimensional vector with the elements being its bin membership -- 1 for a member and 0 for a non-member.
Given the absence of a direct correlation between MIM-Re and MIM-Im, and $(\sigma, \kappa, h)$, we adopt a Binary Cross Entropy loss function. 
In this way, we transform the prediction into a classification problem, where the network output represents the probability for each bin. A good agreement is found between the deep neural network outputs and the ground truth \cite{supp}.

We then apply the trained network to predict the probability distribution of $(\sigma, \kappa)$ for a given set of MIM maps in Fig.~4a. 
The results of $(\sigma, \kappa)$ are shown in Fig. 4c. The obtained $\kappa$ maps indicate the presence of local dielectric disorder in the environmental permittivity, as well as permittivity change of MoSe$_2$ upon back gate doping and optical excitation. The obtained $\sigma$ maps present the spatially dispersive photoconductivity from the RP exciton polaron. These quantified $(\sigma, \kappa)$ maps hence provide a direct measure of the local environment of excitons. 

Following the quantification of $(\sigma, \kappa)$ maps and our understanding of the exciton eigenenergies variation due to charge and electrical environment, as discussed in the last two sections, we formulate the subsequent expression to describe the local exciton eigenenergy $E$ \cite{supp}:
\begin{equation}
    E= E_0 (\kappa) + \beta n - \frac{1}{2}\alpha(\kappa) F^2.
\end{equation}
The first term of the expression, $E_0$, represents the exciton eigenenergy for monolayer MoSe$_2$ under the influence of environmental permittivity, without further considering exciton-electron interactions or electric field effects. The second term represents the interaction between the exciton and free carriers. The variable $n$ corresponds to carrier density, and $\beta$ is a fitting coefficient. The linear relationship is extracted from experimental results, such as Fig. 2f, and the coefficient $\beta$ is derived from the back gate sweeping data. Since this term results from a combination of both bandgap renormalization and the reduction in exciton binding energies with carrier doping, and these effects largely cancel each other out, it is not a main contribution to spatial eigenenergy fluctuation. The third term accounts for the Stark effect resulting from in-plane electric fields with strength $F$. The contribution from out-of-plane electric field is negligible since the in-plane exciton polarizability $\alpha$ is approximately two orders of magnitude larger than the out-of-plane value due to hBN dielectric screening \cite{pedersen2016exciton}. In our measurements, the Stark shift term dominates the spatial variation of exciton eigenenergy $E$.

Subsequently, we determine the exciton eigenenergies $E$ on the 2D grid within the same 1~$\mathrm{\mu}$m $\times$ 1~$\mathrm{\mu}$m region, through fitting and interpolating the MIM hyperspectra. The extracted $E$ is presented in Fig. 4d. With $E$ and the corresponding DNN-predicted $(\sigma,\kappa)$ as inputs for solving Eqn. (1), we generate the predicted electric field distribution, which is illustrated in Fig. 4e. It is evident that the exciton-assisted prediction of the electric field map effectively captures the hotspots in the nanojunction at the graphite boundary (marked as white dotted line), which agrees well in width and amplitude with the finite element simulation results\cite{supp} (Fig. S7). Moreover, it captures the built-in field in other regions hundreds of nanometers away from the junction (i.e., in region II). This result reveals that the photoelectrical properties of a prototypical 2D TMD device are influenced by a combination of factors, such as the Schottky field and fluctuations in conductivity manifested as localized charge puddles. It highlights the critical role of the electric field, which is dispersive on the sub-diffraction limit scale and emphasizes the need to consider the complex interaction of various electrical properties for accurate assessment in 2D materials-based optoelectronic applications. Moreover, this result demonstrates the effectiveness of exciton-assisted electrometry, and the ``all in one" readout of multiple electrical quantities, which is general and can be applied for nano-electrical sensing in a wide variety of quantum systems, demonstrating its potential and versatility in quantum applications.

\section{Conclusions} 

Our work represents a significant advance in the field of nano-imaging for exciton dynamics. The ER-MIM introduces an attempt to drive optoelectronics into the nanoscale, with distinct advantages over traditional methods including superior sensitivity with a low requirement for excitation photon flux, applicability under extreme cryogenic conditions, and high spatial and spectral resolutions. The application of ER-MIM to a gate tunable monolayer MoSe$_2$ device uncovered the intricate interplay between excitons, charge carriers, and built-in electrical fields in a confined dielectric environment at cryogenic temperatures. Employing a neural network, we developed an algorithm for quantitatively determining the local electric field, conductivity, and environmental permittivity. These insights lay the foundations for employing exciton as an all-in-one sensor for local electrical information and mark a leap in advancing sensing modalities for quantum materials and devices.

\section{Methods}\label{sec3}
The full sample fabrication procedure is detailed in ref.\cite{kometter2023hofstadter}. Briefly, graphene, hBN, and MoSe$_2$ were exfoliated from bulk crystals onto Si/SiO$_2$ (285~nm) substrates and assembled in heterostructures using a standard dry-transfer technique with a poly(bisphenol A carbonate) film on a polydimethylsiloxane (PDMS) stamp. MoSe$_2$ was encapsulated between two approximately 30 nm thick hBN flakes. Electron beam lithography and electron beam metal deposition were used to fabricate electrodes for the electrical contacts.

The exciton resonant microwave impedance microscopy (ER-MIM) experiment was conducted with a supercontinuum laser (NKT photonics, FIU-20). The wavelength scans were carried out with a monochromator (Princeton instrument, Acton SP 2300). The linewidth after the monochromator is around 0.5 nm. The laser power at the end of the 100 $\mu$m diameter multimode fiber is $\sim$ 0.1--1~mW throughout the measurements, with the spot radius at the end of the tip being around 0.8 mm. 

\section{Acknowledgements}
Funding: This work was supported by the QSQM, an Energy Frontier Research Center funded by the U.S. Department of Energy (DOE), Office of Science, Basic Energy Sciences (BES), under Award \#DE-SC0021238. Previous development of MIM technique was funded in part by the Gordon and Betty Moore Foundation’s EPiQS Initiative through Grant GBMF4546 to ZXS. B. E. F. acknowledges support from NSF-DMR-2103910 for sample fabrication. K.W. and T.T. acknowledge support from the JSPS KAKENHI (Grant Numbers 20H00354 and 23H02052) and World Premier International Research Center Initiative (WPI), MEXT, Japan. Part of this work was performed at the Stanford Nano Shared Facilities (SNSF), supported by the National Science Foundation under award ECCS-2026822. Z. J. and Z. Z. acknowledge support from the Stanford Science fellowship. Z.J. acknowledge support from the Urbanek-Chodorow fellowship. M.K.L. acknowledges support from the NSF Faculty Early Career Development Program under Grant No. DMR - 2045425. The research is funded in part by a QuantEmX grant from ICAM and the Gordon and Betty Moore Foundation through Grant GBMF9616 to M.K.L. M.K.L. also acknowledges helpful discussion with Xinzhong Chen. Z.Z. acknowledges helpful discussions with Houssam Yassin. Z.J. acknowledges helpful discussions with Tony Heinz and Dung-Hai Lee.

\normalem
\let\oldaddcontentsline\addcontentsline
\renewcommand{\addcontentsline}[3]{}

\bibliographystyle{naturemag}

\let\addcontentsline\oldaddcontentsline
\newpage
\begin{figure}[h!]
 \centering
  \includegraphics[width=0.7\textwidth,trim={5cm 2cm 11cm 1cm},clip]{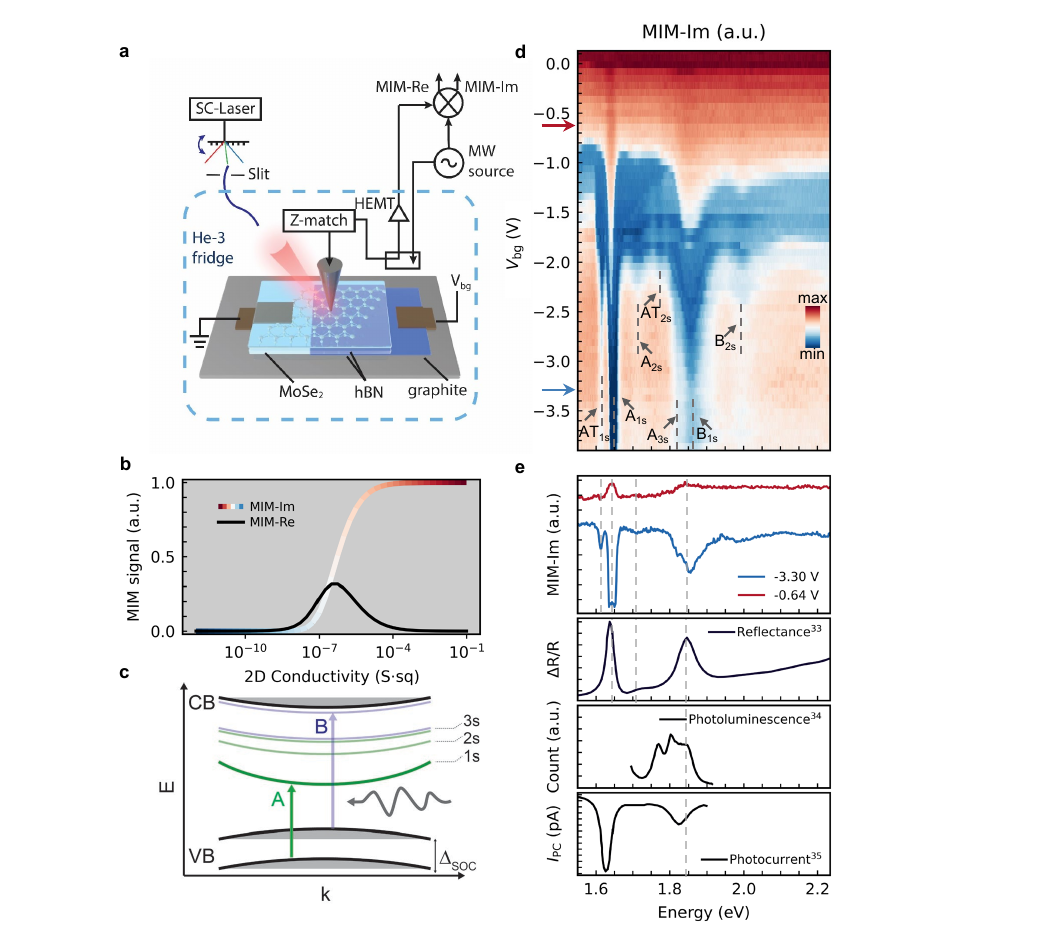}
 \caption{\textbf{Sensing exciton Rydberg states at the nanoscale.} \textbf{a}. The experimental setup of the exciton resonant - microwave impedance microscopy (ER-MIM) measurements. The optical source is a supercontinuum laser. \textbf{b}.  The response curves of MIM-Im and MIM-Re as a function of the sheet conductivity of MoSe$_2$. The darker blue color indicates a conductivity decrease relative to neighboring regions, and a deeper red a conductivity increase. \textbf{c}. A schematic showing relevant energy levels and optical transitions forming 1s, 2s, and 3s Rydberg states of A and B excitons in monolayer MoSe$_2$.  \textbf{d}. ER-MIM-Im signal as a function of back gate voltage $V_\text{bg}$ and optical excitation energy. The blue and red arrows on the y axis correspond to two representative doping levels. \textbf{e}. Comparison of ER-MIM results (at the two doping levels marked in \textbf{d}) with spectra from three widely used far-field optical techniques: optical reflectance\cite{arora2015exciton}, photoluminescence\cite{liu2021exciton}, and photocurrent\cite{vaquero2022low} measurements.}
  \label{fig:1}
 \end{figure}
 \newpage
 \begin{figure}[h!]
 \centering
  \includegraphics[width=0.7\textwidth,trim={7cm 6cm 11cm 1cm},clip]{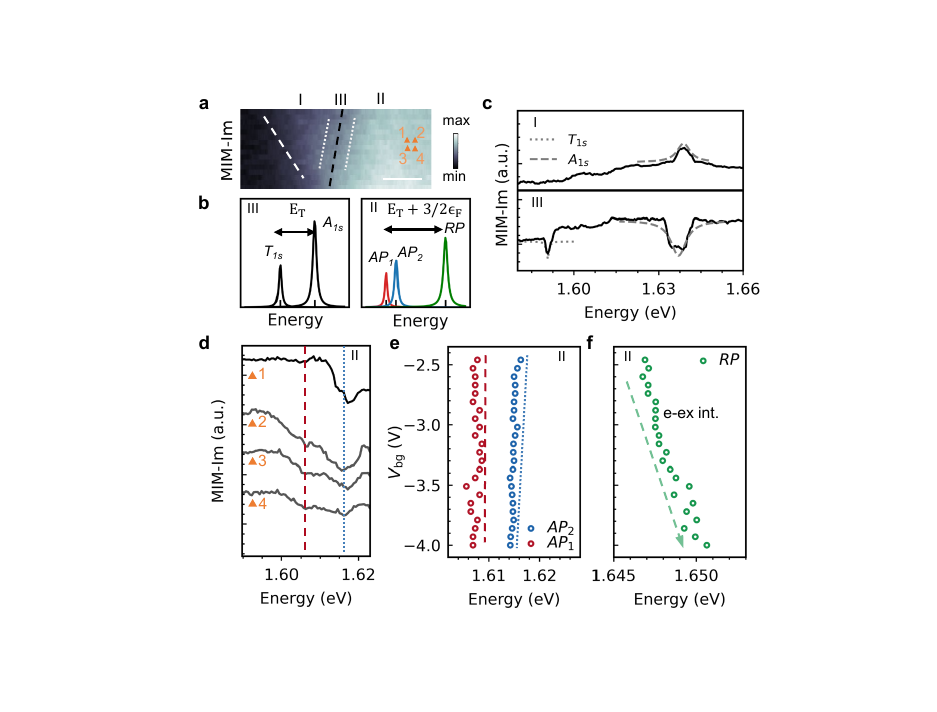}
 \caption{ \textbf{The interplay between exciton behaviors with local charge environment.} \textbf{a}. A MIM map showing the MoSe$_2$ sample (flake boundary outlined by a white dashed line) partially on top of a graphite back gate (boundary outlined by a black dashed line). The device can be divided into three regions, I (outside of graphite), II (above graphite), and III (depletion region, with boundaries outlined by two white dotted lines). Scale bar: 500~nm \textbf{b}. Schematics of the exciton and trion spectra (left), and the exciton-polaron spectra (right). \textbf{c}. The ER-MIM-Im spectrum measured in region I (top panel) and III (bottom panel). The grey dashed and dotted lines are fittings of $A_\text{1s}$ exciton and $T_\text{1s}$ trion. \textbf{d}. ER-MIM-Im spectra measured at four locations in region II with a fixed gate voltage. The locations are marked by four orange triangles in \textbf{a}. \textbf{e}. The evolution of two eigenenergies of the A exciton: the red dashed and blue dotted lines are the two attractive branches AP$_1$ and AP$_2$, respectively. \textbf{f}. The evolution of the repulsive polaron branch of the A exciton, as shown by the green dashed line. `e-ex. int' stands for exciton-electron interactions.}

  \label{fig:2}
 \end{figure}

\newpage
 \begin{figure}[h!]
 \centering
  \includegraphics[width=1\textwidth,trim={3.8cm 0 9cm 0},clip]{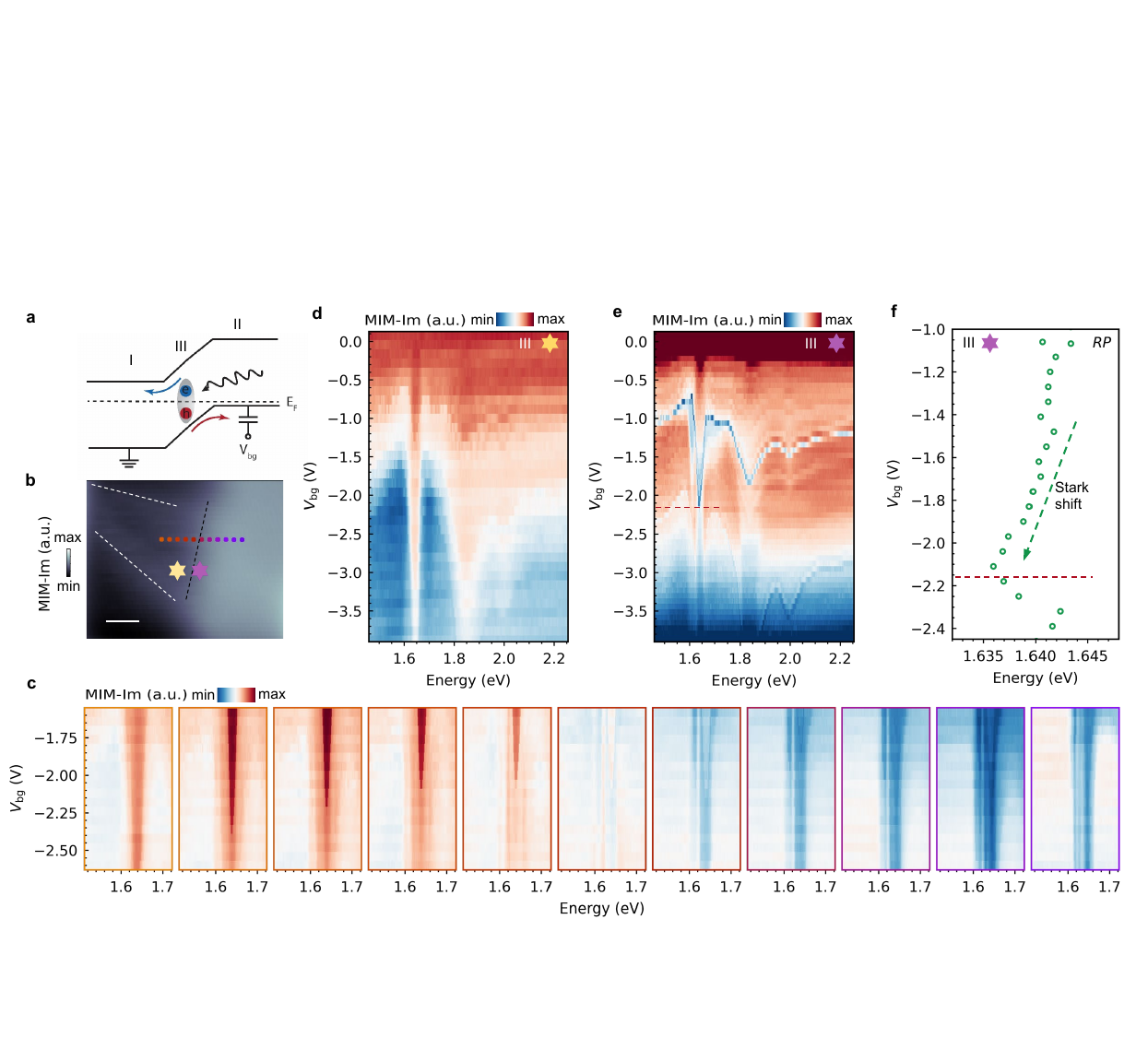}
 \caption{\textbf{The interplay between exciton behaviors and electric environment.} \textbf{a}. A schematic of the $p$-$n$ junction formed through V$_{bg}$ modulation in region II, and the depletion region formation in III. \textbf{b}. MIM-Im image showing the sample geometry, with colored circles and stars marking different locations of further measurements. (The same region as Fig. 2a but at a larger scale, with the MoSe$_2$ flake boundary outlined by white dashed lines, and the graphite boundary outlined by a black dashed line.) Scale bar: 500~nm. \textbf{c}. A series of ER-MIM-Im measurements at the locations specified by the colored circles in \textbf{b}. The colors of the boxes are in correspondence with the color of the circles (plots ordered in the sequence from left to right). \textbf{d}. ER-MIM-Im measured at the yellow star in \textbf{b}. \textbf{e}. ER-MIM-Im measured at the purple star in \textbf{b}. \textbf{f}. The extracted repulsive polaron eigenenergy as a function of back gate voltage. The red dashed lines in \textbf{e} and \textbf{f} mark the $V_\text{bg}$ at which the RP exciton energy has a sudden blueshift.}
 \end{figure}
\newpage
  \begin{figure}[h!]
 \centering
  \includegraphics[width=1\textwidth,trim={4cm 5cm 10cm 0},clip]{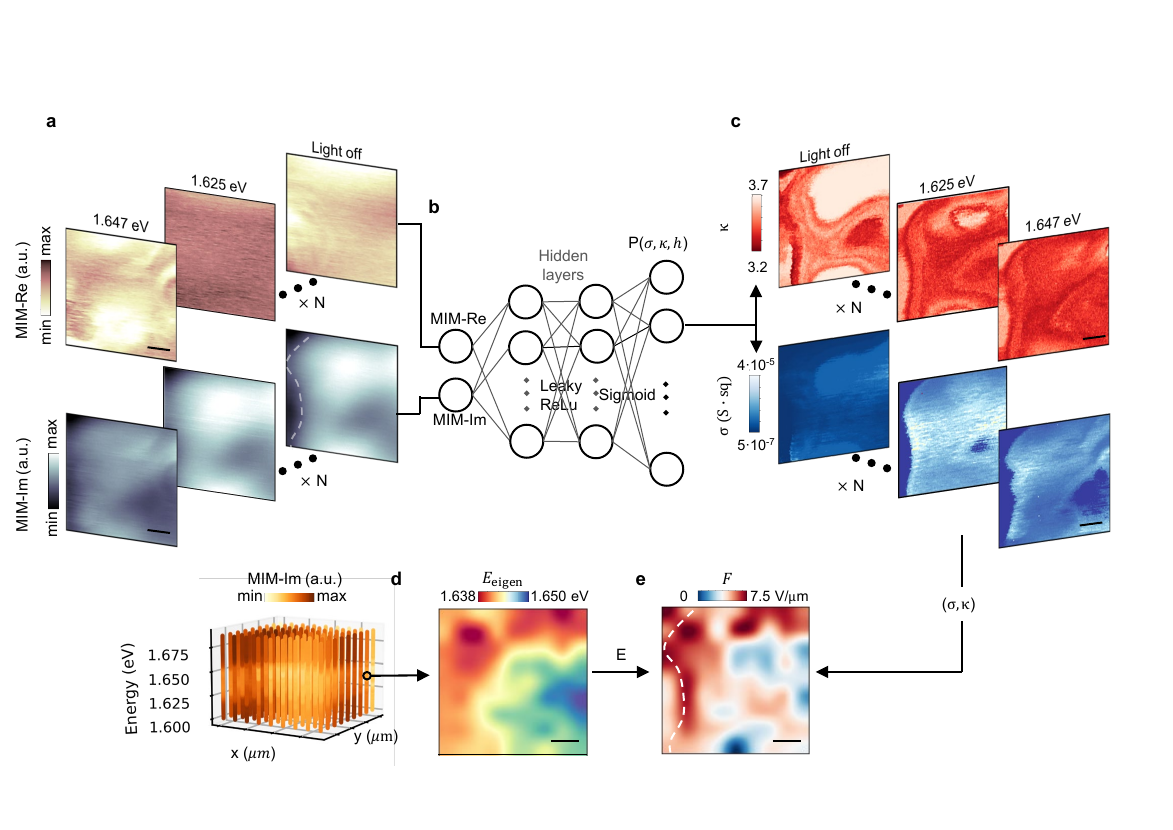}
 \caption{ \textbf{Holistic electrical sensing with excitons.} \textbf{a}. Representative sets of MIM-Re and MIM-Im images measured with light off, 1.625 eV (away from exciton resonance), and 1.647 eV (close to A$_{1s}$ repulsive exciton-polaron resonance), in a 1~$\mathrm{\mu m} \times 1$~$\mathrm{\mu m}$ region. The white dashed line denotes the graphite boundary. Scale bar: 200~nm. \textbf{b}. The deep neural network architecture. \textbf{c}. The deep neural network training result of permittivity $\kappa$, conductivity $\sigma$ for the three columns of MIM data in \textbf{a}. Scale bar: 200~nm. \textbf{d}. Left: The ER-MIM-Im hyperspectra as a function of excitation energy and spatial location collected on a 9$\times$9 grid. Right: The interpolated exciton eigenenergy extracted from the hyperspectra. Scale bar: 200~nm. \textbf{e}. The interpolated in-plane electric field strength prediction based on Eq.~(1), and the input from $(\sigma,\kappa)$ in \textbf{c} and $E$ in \textbf{d}, as indicated by the black flow arrows. The white dotted line denotes for the graphite boundary. Scale bar: 200~nm.}
 \end{figure}


\clearpage

\setcounter{equation}{0}
\setcounter{figure}{0}
\setcounter{table}{0}
\setcounter{page}{1}
\setcounter{section}{0}
\makeatletter
\renewcommand{\theequation}{S\arabic{equation}}
\renewcommand{\thefigure}{S\arabic{figure}}
\renewcommand{\thepage}{S\arabic{page}}

\begin{figure}[h!]
 \centering
  \includegraphics[width=0.8\textwidth,trim={7cm 9cm 11cm 1cm},clip]{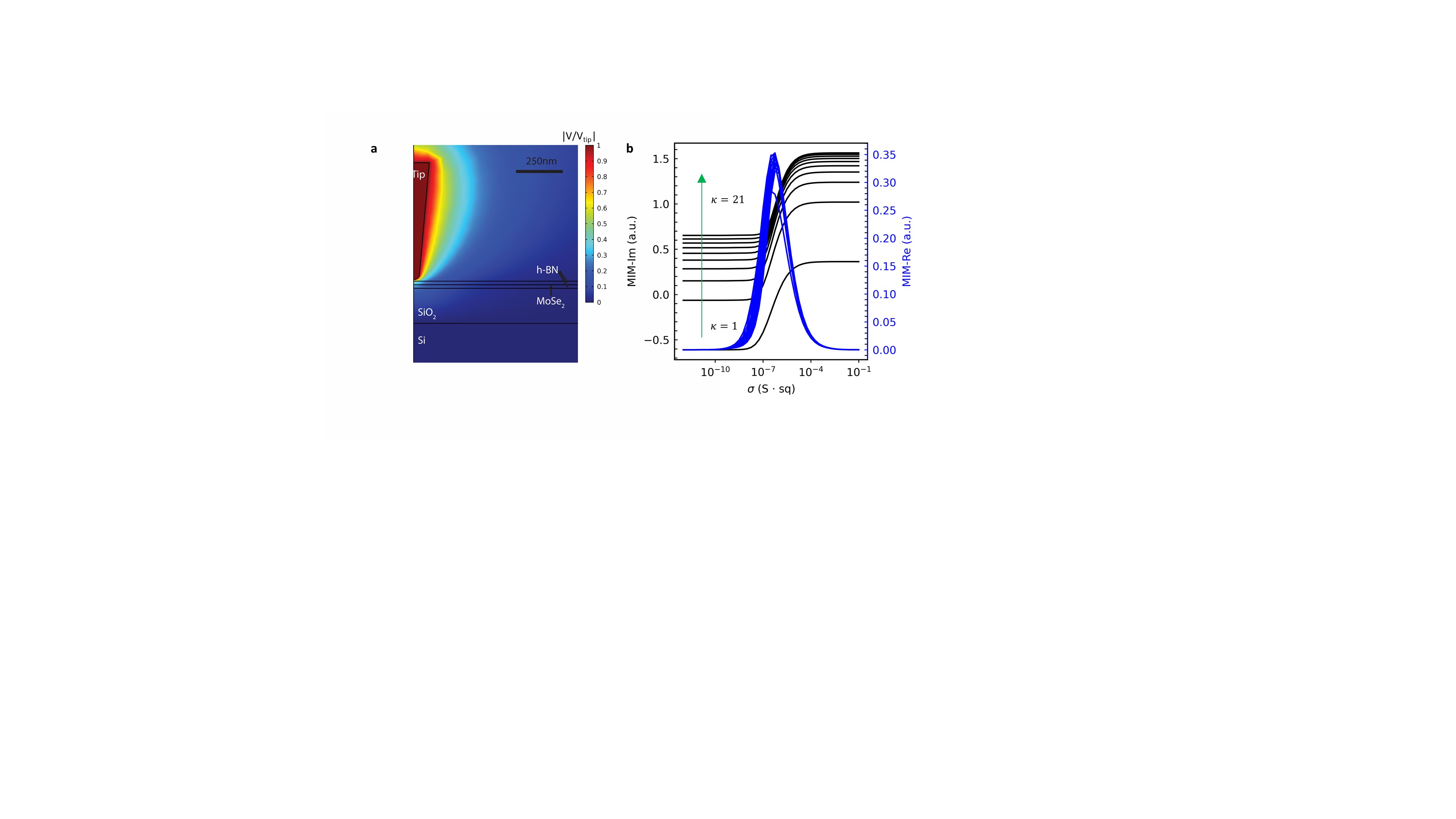}
 \caption{\textbf{a}. Tip and sample device geometry, and finite element calculations of the quasi-static potential distribution ($V$). The colorplot is $|V/V_{\text{tip}}|$, with $\sigma=10^{-12}S\cdot sq$ and $\kappa=3$. \textbf{b}. The simulated MIM-Im, MIM-Re response curves as a function of the sheet conductance $\sigma$ of MoSe$_2$ and relative permittivity $\kappa$ (from 1 to 21, with a fixed increment of 2) of the slab. }
  \label{fig:comsol}
 \end{figure}

 \begin{figure}[h!]
 \centering
  \includegraphics[width=0.6\textwidth,trim={8cm 8.5cm 10cm 1cm},clip]{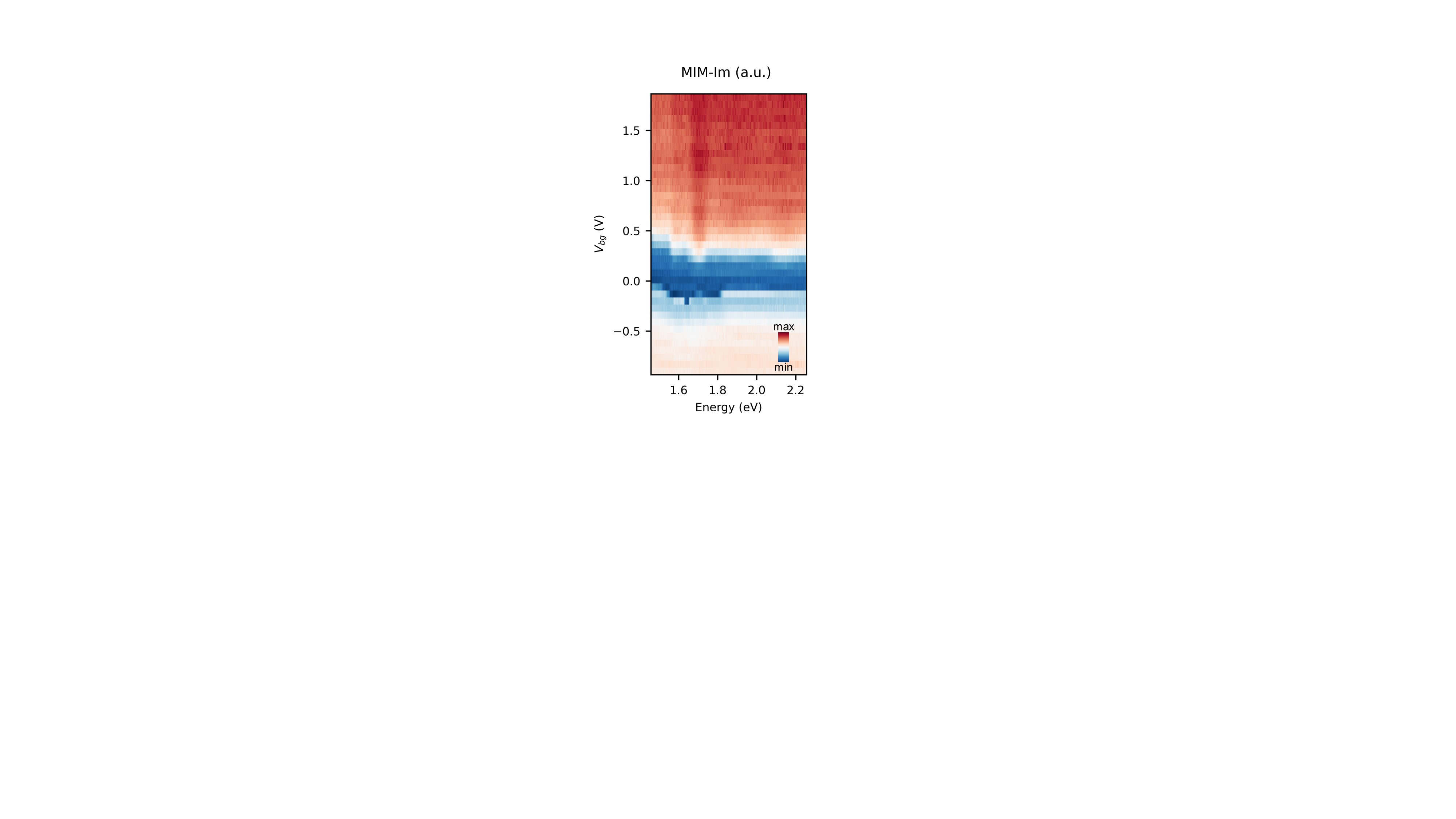}
 \caption{MIM-Im signal measured on monolayer WSe$_2$, as a function of back gate voltage $V_{\text{bg}}$ and optical excitation energy.}
  \label{fig:wse2}
 \end{figure}

  \begin{figure}[h!]
 \centering
  \includegraphics[width=0.8\textwidth]{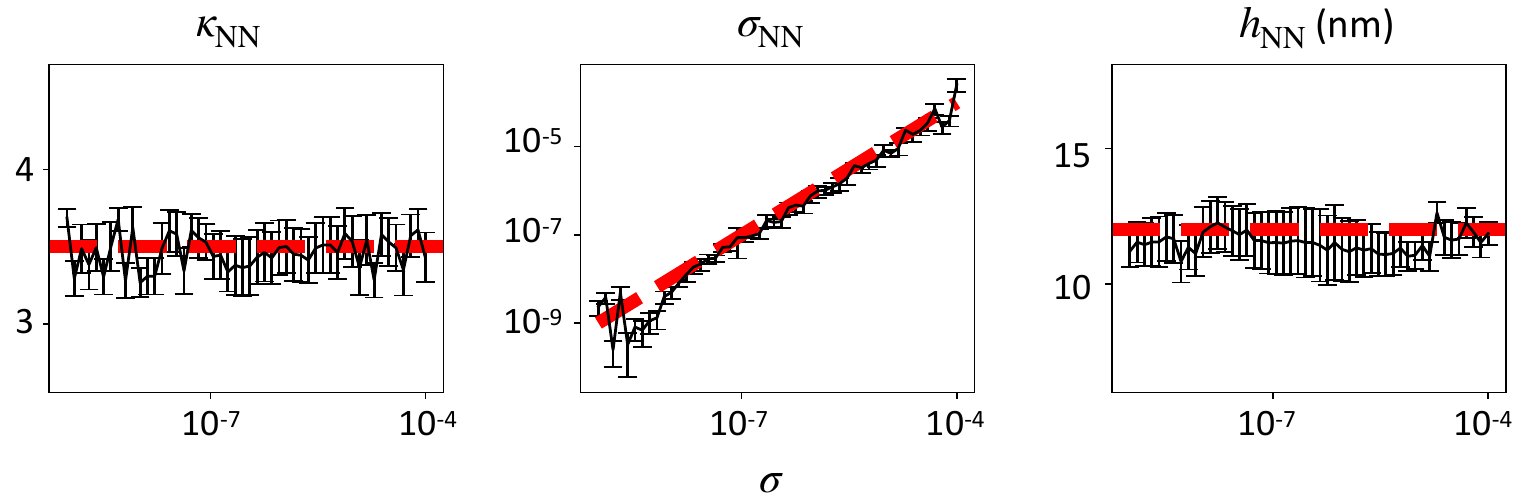}
 \caption{Neural-network predicted $\epsilon$, $\sigma$, and $h$ as functions of the ground truth conductivity, averaged over 100 samples. The red dashed lines represent the ground truth. Input data is obtained from interpolating simulations with random noise of amplitude 4\%, while varying conductivity $\sigma$ and keeping permittivity and sample-tip distance fixed at $\epsilon = 3.5$ and $h = 12 \, \mathrm{nm}$, respectively. }
  \label{fig:test}
 \end{figure}

 \begin{figure}[h!]
 \centering
  \includegraphics[width=1\textwidth,trim={5cm 5cm 5cm 1cm},clip]{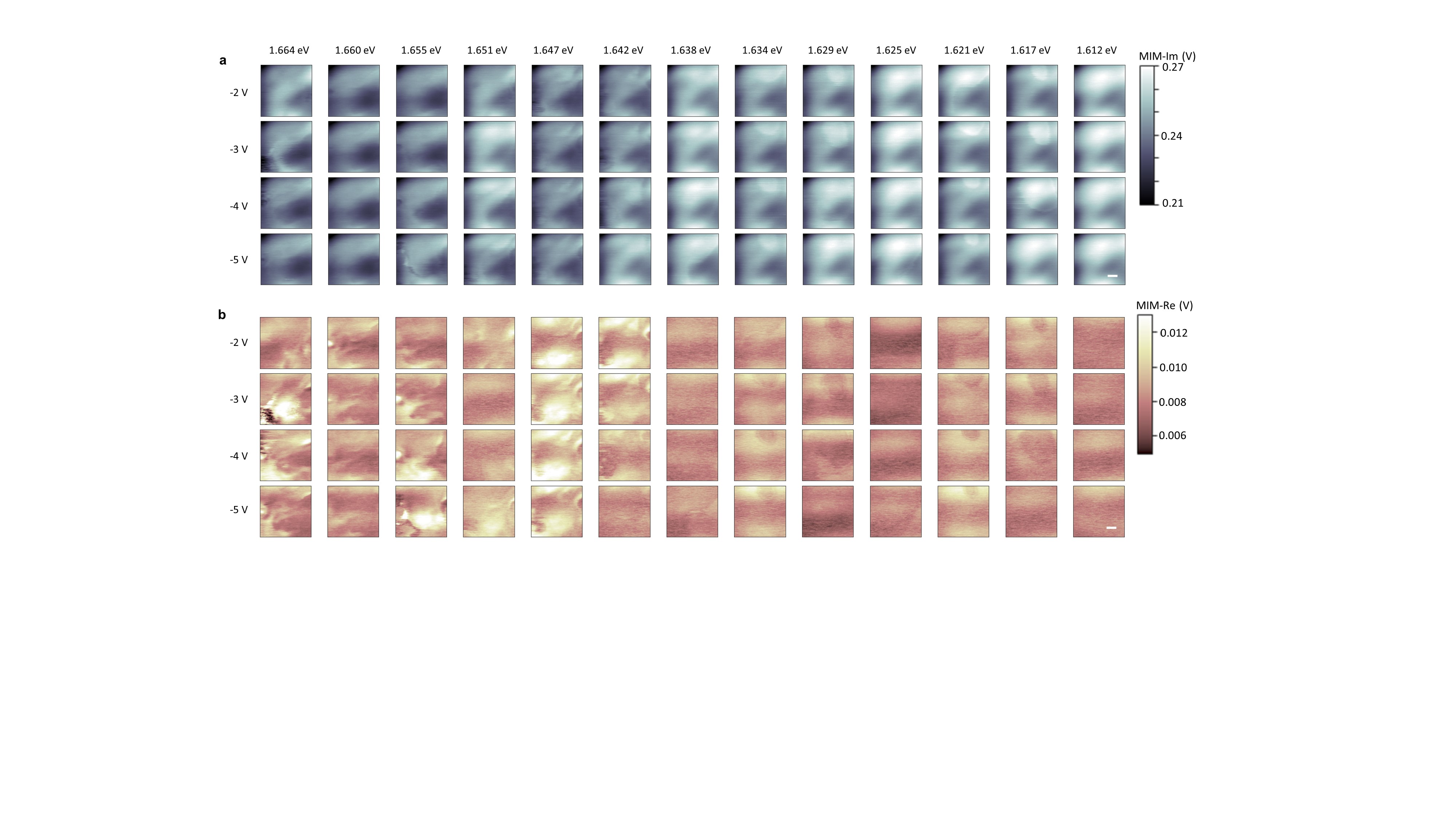}
 \caption{\textbf{a}. MIM-Im and \textbf{b}. MIM-Re images measured at a set of optical excitation energies and back gate voltages in a 1~$\mathrm{\mu m} \times 1$~$\mathrm{\mu m}$ region. Scale bar: 200~nm. }
  \label{fig:mim_orig}
 \end{figure}

  \begin{figure}[h!]
 \centering
  \includegraphics[width=1\textwidth,trim={7.25cm 5cm 2.75cm 1cm},clip]{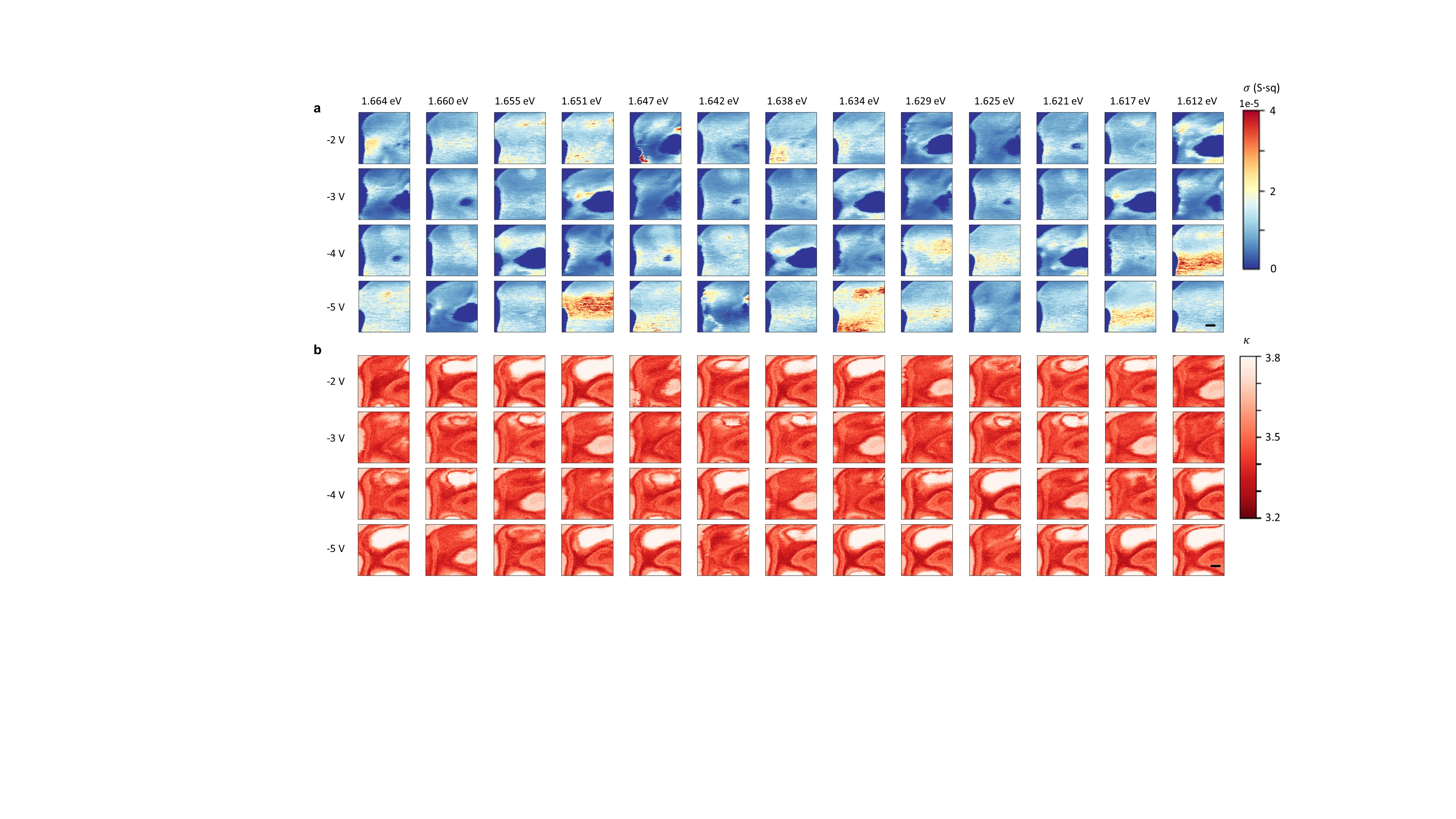}
 \caption{The neural network predictions of \textbf{a}. $\sigma$ in units of S$\cdot$sq and \textbf{b}. $\kappa$, from the MIM images measured at a set of optical excitation energies and back gate voltages in a 1~$\mathrm{\mu m} \times 1$~$\mathrm{\mu m}$ region. Scale bar: 200~nm. }
  \label{fig:sigma_pred}
 \end{figure}

  \begin{figure}[h!]
 \centering
  \includegraphics[width=1\textwidth,trim={6.5cm 4cm 3.5cm 1cm},clip]{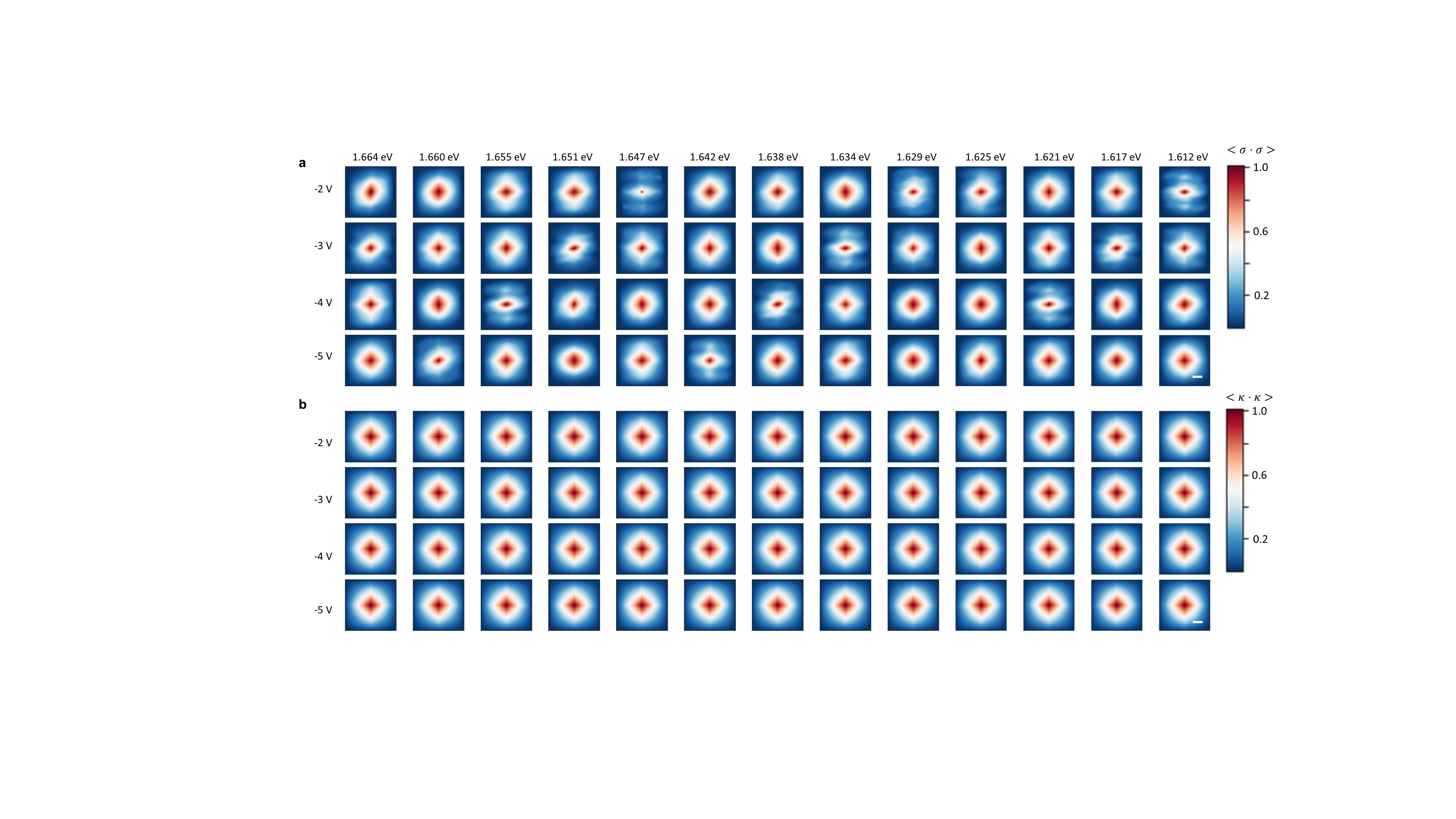}
 \caption{Autocorrelation images of the neural network predicted \textbf{a}. $\sigma$ and \textbf{b}. $\kappa$, from the MIM images measured at a set of optical excitation energies and back gate voltages in a 1~$\mathrm{\mu m} \times 1$~$\mathrm{\mu m}$ region. Scale bar: 200~nm. }
  \label{fig:corr}
 \end{figure}

\begin{figure}[!ht]
\centering
\includegraphics[width=0.3\textwidth,trim={4cm 3cm 23cm 1.5cm},clip]{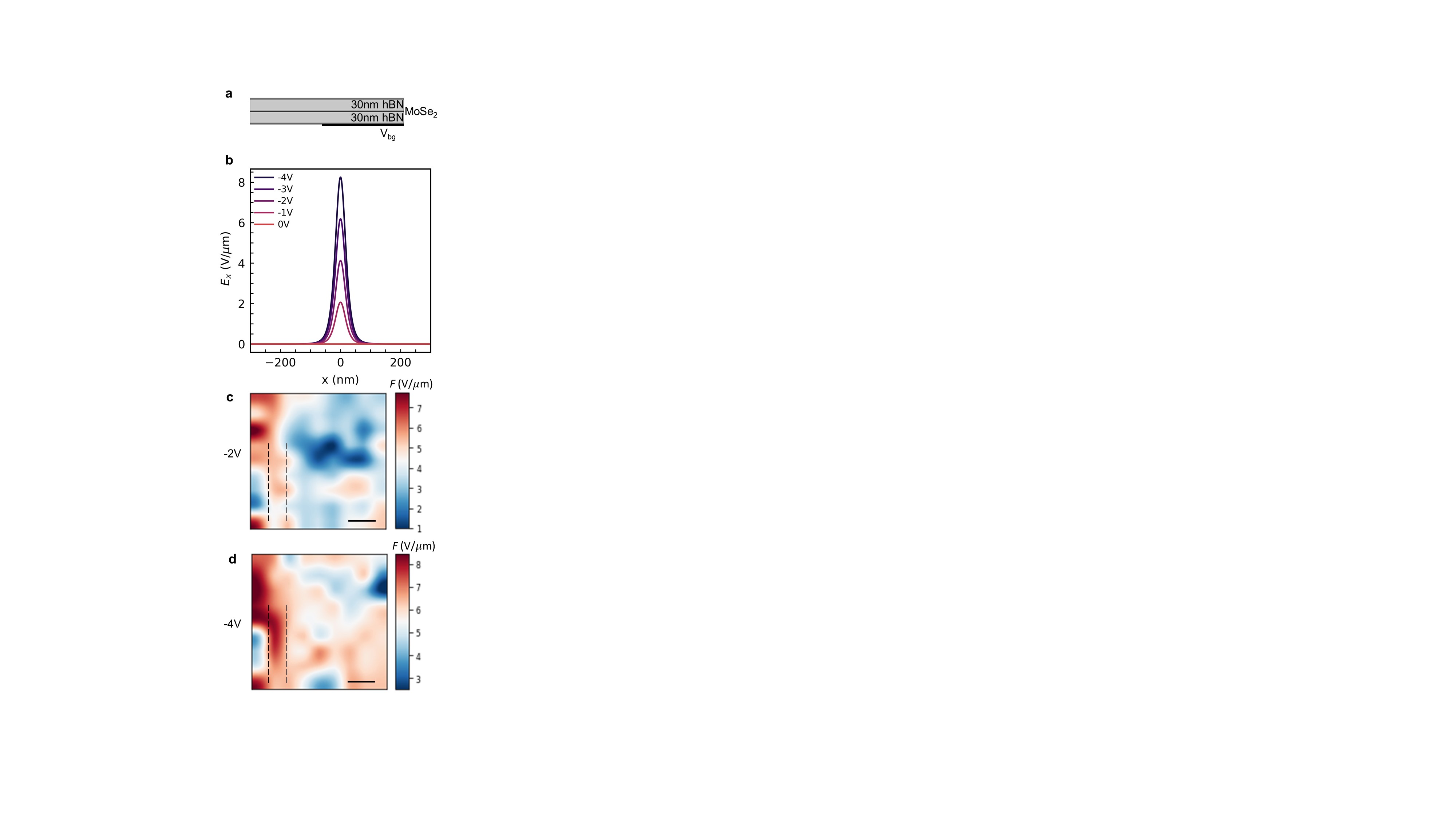}
\caption{\textbf{a}. Geometry for the finite element simulation of the in-plane electric field. \textbf{b}. The Comsol simulated electric field distribution with various back gate voltages applied to the right half of the device. \textbf{C, D}. The predicted, interpolated electric field distribution in the 1 $\mu$m by 1$\mu$m region, at \textbf{c}. $V_\text{bg}= -2\,\mathrm{V}$ and \textbf{d}. $V_\text{bg}= -4\,\mathrm{V}$. The black dashed lines mark the depletion region centered at the graphite boundary, in accordance with \textbf{b}. Scale bar: 200 nm.}
\label{fig:Efield}
\end{figure}

\onecolumngrid
\clearpage
\begin{center}
\textbf{\large Supplementary information for ``Harnessing excitons at the nanoscale - photoelectrical platform for quantitative sensing and imaging'' }\\[5pt]
\begin{quote}
 {\small 
}
\end{quote}
\end{center}
\section{Finite-element analysis of the MIM signals}
Finite-element analysis (FEA) of the MIM response to 2D sheet conductance is shown in Fig.~\ref{fig:comsol}. For the tip-sample geometry in Fig.~\ref{fig:comsol}a, the tip radius is chosen to be 50 nm, and the distance above the sample surface 10 nm. The sample parameters are: The monolayer MoSe$_2$ is 1 nm in thickness, both top and bottom hBN layers are 30 nm in thickness. The hBN - MoSe$_2$ - hBN slab is approximated with a uniform relative permittivity $\kappa$, since hBN and MoSe$_2$ have similar quasi-static dielectric constants \cite{laturia2018dielectric,raja2019dielectric}. The SiO$_2$ substrate is 300 nm thick with $\kappa_{\text{SiO}_2}$ = 3.9. The simulated MIM-Im, MIM-Re response curves as a function of the sheet conductance of $\sigma_{\text{MoSe}_2}$ and relative permittivity $\kappa$ of the slab are shown in Fig. \ref{fig:comsol}b. Both simulations with and without the graphite layer were conducted, and they differ by a scaling factor, so the simulation result is shown in arbitrary units.

\section{Possible origins of the ER-MIM response}
To understand the origins of the ER-MIM response, we discuss different photoconductivity generation mechanisms in a 2D TMD device upon exciton formation and dissociation. 

1. Microwave induced dissociation of exctions: With an input power of $\sim$0.2 $\mu$W, and considering the gain offered by the half-wave impedance-match section, the GHz voltage at the tip is on the order of 0.01 V \cite{shan2022universal}. Then, with the tip radius parameter of $\sim$ 50 nm, and considering the dielectric constant of hBN, we estimate that the typical electric field at the sample is less than 0.1  V/$\mu$m. This is much smaller than the gate voltage applied to the monolayer and hence not considered a major process \cite{massicotte2018dissociation}. 

2. Photogating effect: The photogating effect can be excited through direct photoactivation of midgap charged defects in hBN \cite{ju2014photoinduced}. But the observation that the photoconductivity has dips near the exciton absorption resonances implies that it depends on the exciton population in MoSe$_2$, instead of the hBN defect population itself, which rules out this mechanism. For the other type of photogating effect proposed to be happening at interfaces, our sandwiched device geometry would have two interfaces, making the segregation of electrons or holes on one specific interface relatively unlikely.

3. Auger effect: This effect is largely dependent on the band alignment of the TMD monolayer and hBN. To prove this argument, we also conduced measurements on the monolayer WSe$_2$ part of the device (Fig. \ref{fig:wse2}), where it shows less prominent ER-MIM-Im decrease at excitonic resonances with similar measurement conditions as in Fig. 1d. This is in accordance with the DFT calculation which shows a larger valence band misalignment between WSe$_2$ and hBN and hence less tunneling \cite{sushko2020asymmetric}. 

4. Photoconductivity: The exposure of TMD-based devices to light can generate photo-carriers resulting in an enhanced conductivity. The dissociation of electron hole pairs is fostered by the existence of defects, including hole traps in the MoSe$_2$. A similar enhancement of photoconductivity near excitonic excitations were measured on monolayer WSe$_2$ (Fig. \ref{fig:wse2}, vertical stripes in dark red color). 

\section{Details of the machine learning algorithm}

Our goal is to predict the local environment ($\sigma$, $\kappa$, $h$) for a given set of MIM real and imaginary measurements, MIM-Re and MIM-Im. However, this mapping is not bijective, as multiple sets of ($\sigma$, $\kappa$, $h$) can correspond to the same (MIM-Re, MIM-Im). Therefore, instead of predicting a fixed ($\sigma$, $\kappa$, $h$), we assign a probability to each set of ($\sigma$, $\kappa$, $h$) for a given measurement.
Figure 4 in the main text shows the feedforward network architecture used for predicting local properties. The network takes input values of MIM-Re and MIM-Im, either from experiments or simulations and predicts the probability distribution for discretized $(\sigma, \kappa, h)$ values.

For training the network, we start by simulating MIM-Re and MIM-Im for various sets of $(\sigma, \kappa, h)$. We then interpolate this simulation data across a physically relevant range of $(\sigma, \kappa, h)$ values: $\sigma$ from $1\times10^{-11}$ to $6\times 10^{-3}$ S$\cdot$sq, $\kappa$ between 3.2 and 3.9~\cite{laturia2018dielectric}, and $h$ from 6 to 18~nm, with even spacing in each dimension. This amounts to a total of 64,000 sets of data. We further divide the data into bins: 32 bins for $\sigma$, 8 bins for $\kappa$, and 15 bins for $h$, resulting in 3,840 bins in total. Each data point is then encoded into a 3,840-dimensional one-hot encoding vector based on its bin membership. During training, we minimize the Binary Cross Entropy loss, transforming the prediction into a classification problem, where the network output is a 3,840-dimensional vector representing the probabilities for each bin.

The network architecture consists of 2 hidden layers with 100 nodes each. We use the Adam optimizer with a constant learning rate of $0.01$ and set $\beta_1$ and $\beta_2$ to 0.999 and 0.9999, respectively. We train until the variance of the loss function over the last 200 steps falls below $1\times 10^{-5}$. Data is randomly split into 80\% for training, 10\% for validation, and 10\% for testing.
Figure~\ref{fig:test} shows the neural network predictions for interpolated simulation data with varying conductivity and random noise amplitude 4\%. The predictions and the ground truth agree within errorbars for all values of conductivity.

We proceed to apply the trained network to predict the probability distribution of $(\sigma, \kappa)$ for a given set of MIM maps in Fig. \ref{fig:mim_orig}. 
All the measurements are of the same region of the sample so the sample topography is the same for all the different data sets. 
To fix the height distribution, we first condition one set of MIM measurements to be similar to an initial guess distribution, consisting of a jump in height with a smooth edge at the junction as well as a two-dimensional Gaussian distribution on the right representing a bubble. 
To achieve this, we calculate the distance, $D$, of each bin from $(\epsilon_0, h_0)$ as $D_j = |h_j - h_0|$ and weight the predicted probability distribution with $\exp(-\alpha D_j)$, where $\alpha$ denotes the constraint strength (with $\alpha = 2$ in this case). 
In addition, we impose a smoothness condition by encouraging the neighboring pixels to be similar. 
For all predictions, we average over 20 samples per pixel. 
The results of $(\sigma, \kappa)$ are shown in the first two columns of Fig.~\ref{fig:sigma_pred}. 
The prediction for $\kappa$ and $\sigma$ share some similarities, due to the presence of local fluctuations in environmental permittivity, or dielectric disorder, as well as charge screening effects. Furthermore, when comparing among different sets of $(\sigma, \kappa)$, the contrast is observed between excitation energy close or further away from the excitonic resonances. These quantified $(\sigma, \kappa)$ maps hence provide a direct measure of the exciton-induced photoconductivity and the impact of the dielectric environment. 

Figure \ref{fig:corr} shows the autocorrelated images of $\sigma$ and $\kappa$. Notably, while the correlation of relative permittivity remains relativity consistent, the normalized autocorrelation of conductivity $<\mathbf{\sigma} \cdot \mathbf{\sigma}>$ has a decrease when the RP branch of excitons is on resonance. This decrease is more prominent along the vertical direction, perpendicular to the direction of the $p$-$n$ junction. It provides a signature of exciton excitation induced conductivity variation, and supports the self-consistency of the approach.

\section{Details of electric field prediction}
\subsection{Benchmarking using Comsol simulations}
To cross check the electric field sensing result, we performed finite element simulations to predict the electric field distribution near the junction. In this simulation, we assumed zero temperature, and modelled the MoSe$_2$ monolayer as a charge sheet with density \cite{thureja2022electrically},
\begin{equation}
    \begin{aligned}
    n(x)&=n_e(x)+n_h(x)\\
    &=-eD(E)(E_{\text{F}}-E_{\text{C}})+eD(E)(E_{\text{V}}-E_{\text{F}})
    \end{aligned}
    \label{som2}
\end{equation}

where $n_e(x)$ and $n_h(x)$ are the electron and hole charge densities, respectively. They are determined by the 2D density of states $D(E)=2m^*/(2\pi\hbar^2)$ ($m^*$ is the effective mass). $E_{\text{C}}$ and $E_{\text{V}}$ are the conduction and valence band edge energies, respectively, which are dependent of the local electrostatic potential. $E_{\text{F}}$ is the Fermi level determined by the alignment of the contact work function with respect to the band edges.

The simulated device geometry is depicted in Fig. \ref{fig:Efield}. The sheet of charge is encapsulated by two 30~nm thick hBN slabs and contacted by an ohmic electrode. Furthermore, we include a gate with only partial coverage. The voltage on the back gate is $-$4~V in our simulations. The material parameters used for this calculation are as follows: MoSe$_2$ band gap $E_{\text{b}} = 1.4$~eV, Fermi level offset relative to the valence band edge at zero potential $E_{\text{F}}-E_{\text{V}}(V = 0) = 0.9$~eV, electron effective mass $m_n=0.56 m_e$, hole effective mass $m_p=0.59m_e$ \cite{kormanyos2015k}, hBN dielectric constant $\kappa=3.76$.

In this way, we obtain the in-plane electric field distribution depicted in Fig. \ref{fig:Efield}b at various back gate voltages on the right half of the device. We then compare the electric field prediction results shown in Fig.~\ref{fig:Efield} c,d, which are measured at $V_{\text{g}}= -2\,\mathrm{V}$ and $-4\,\mathrm{V}$, respectively, with the simulation result. It shows agreement of both amplitude and width of the depletion region near the boundary. 

The input of exciton eigenenergies is critical to obtain the correct electric field distribution, as direct finite element simulations from $\sigma$ and $\kappa$ has limitations. The limitations involve several aspects: 1) The built-in field formed by conductivity fluctuations is hard to capture without accurate measurement of carrier mobilities. 2) The sophistication of semiconductor properties of 2D TMD materials at cryogenic temperatures, including defect levels and partial ionizations is hard to quantify. 3) The optical excitation and associated quasi-equilibrium carrier dynamics are hard to quantify.

\subsection{Details of excitonic electrometry and outlook}
The $\beta$ coefficient is derived from fitting the gate-dependent excitonic resonance energy, and the value of $\beta=-2.55\times10^{-3}\,\mathrm{eV /V}$ at $\kappa = 3.5$ is adopted. The spatial variation of $\kappa$ is considered in calculations.
The exciton polarizability is a function of the
effective exciton Bohr radius ($d$) and effective exciton temperature, in a simplified formula of $\alpha_{2D}=e^2d^2/2k_BT$~\cite{xiao2023coulomb}. At our measurement temperature of 1.5 K, there has not been a consensus on the value of polarizability. Our fitted value from experiment, approximately $\alpha = 3.6 \times 10^{-16} \,\mathrm{eV m^2/V} $ (varying with local permittivity), is larger than the theoretically calculated value of $ \sim 2.5 \times 10^{-17} \mathrm{eV m^2/V}$~ \cite{pedersen2016exciton}, while in better agreement with some recent measurements \cite{lu2023giant}. This is due to the low measurement temperature, where the $1/T$ dependence gives rise to much larger polarizability, and also the influence from the dielectric environment. The dependence of $\alpha$ on dielectric constant $\kappa$ is adopted from the theoretically calculated functional form $\alpha = (27.3766-1.2964\kappa+2.086\kappa^2)\times10^{-17}\, \mathrm{eV (m/V)^2}$ ~\cite{pedersen2016exciton} and experimental fitting. The exciton binding energy change with dielectric screening is also incorporated in the form of $\delta E_0 (\kappa)= 0.002\kappa^2-0.032\kappa$\cite{pedersen2016exciton}.

When assuming the spectral linewidth of 0.05\,meV, with these numbers, the exciton sensor has the smallest detectable field of sub 1~$\mathrm{V \cdot \mu m^{-1}}$. The distance between an exciton sensor and a sample can be brought to very closer, due to the 2D nature of TMD monolayers. This sensitivity can be further improved by: 1) Using excitons in other TMD materials, for example in monolayer MoTe$_2$, with almost invariant eigenenergies under changing carrier densities. It can reduce the influence of other terms in main text Eqn.~(1) \cite{biswas2023rydberg}.  2) Utilizing excited Rydberg states of excitons under higher fluence photon excitation, which will yield larger polarizability ($\alpha$). The enhancement has been predicted to follow a scaling law, where $\alpha$ increases with the principal quantum number ($n$) as $\alpha \propto n^7$.\cite{heckotter2017scaling} 3) Utilizing exciton states with distinct in-plane and out-of-plane polarizabilities, such as interlayer excitons in a heterobilayer device. This will assist in achieving a vector electric field readout.
\newpage

\end{document}